# Vaidman's nested MZI – again

B. E. Y. Svensson

*Theoretical High Energy Physics,*

*Department of Astronomy and Theoretical Physics,*

*Lund University, Sölvegatan 14, SE-22362 Lund, Sweden*

E-mail: Bengt_E_Y.Svensson@thep.lu.se

**Abstract**

I provide additional arguments for refuting Vaidman's weak value analysis of the path of a particle in a nested Mach-Zehnder interferometer. The argument uses sequential weak values to supplement my previous analysis according to which even weak measurements disturb the system to such an extent that projector weak values cannot be considered "representative" for the undisturbed system.

In [1] (with many follow-up articles, one of the latest being [2]), Vaidman proposed and employed a criterion for the past of a quantum particle: a non-vanishing weak value of the projector $\Pi_a =$ | $a ><a$ | onto a "path" | $a >$ for a quantum particle reveals that the particle has a non-zero probability of taking that path in its evolution from the initial , "preselected" state | $in >$ to the final, "postselected" state | $f >$. This criterion is basically sound, and lies behind several applications of the weak value approach to specific systems (see, e.g., [3 and [4] and the many references cited therein). It could, however, lead to non-valid conclusions, of which, in my opinion, the application to the nested Mach-Zehnder interferometers (MZIs) setup is an example.

Indeed, the nested MZIs setup (see the figure, also for my notation) is an almost paradigmatic example of the application of Vaidman's criterion. In it, he [1,2] considers the weak values of the projectors onto the different arms of the set-up. The result is astonishing: with prefect 50-50 beamsplitters in the setup, and with the postselected state | $f >$ chosen as the state reaching the detector $D_2$, one finds that the weak values of the projectors $\Pi_D$ and $\Pi_E$ onto the arms $D$ and $E$, respectively, are zero, while the weak values of the projectors $\Pi_B$ and $\Pi_C$ onto the arms $B$ and $C$, respectively, are non-vanishing. Vaidman's conclusion is that the particle appears in the arms $B$ and $C$ without passing the arms $D$ or $E$.

Many physicists find this appearance and disappearance of a particle very odd, not to say impossible, even for a quantum particle. Ever since Vaidman first put forward his proposal [1], there has therefore been an ongoing debate between Vaidman and his opponents concerning the correctness of Vaidman's arguments ( see, *e.g.*, [4] and references therein). All of these objections have been contested by Vaidman (see , *e.g.*, [2] for one of his latest rebuttals).

I am one of the opponents to Vaidman's result. In [5] (see also [6,7] ), I argued that for the nested MZI set-up you cannot take the weak value $(\Pi_B)_w$ as a *bona fide* property of the *undisturbed* system. In more detail, my argument is that even a weak measurement of $\Pi_B$, *i.e.*, testing for the presence of the particle in arm *B* by weak measurement, so to speak selects the arm *B* at the expense of the arm *C*. By this I mean that the weak measurement destroys the destructive interference in the beamsplitter *BS3*, so essential for explaining why the undisturbed system does not allow the particle to reach arm *E*. Thus, even a weak measurement of $\Pi_B$ results in allowing the particle to reach arm *E* – and ultimately the detector $D_2$ – resulting in a non-vanishing weak value $(\Pi_B)_w$. Since the non-vanishing of $(\Pi_B)_w$ depends crucially on the effect of this disturbance, the situation with a weak measurement of $\Pi_B$ does not represent the undisturbed system and cannot be used to characterize it [5].

In this short note, I take this argument one step further. I want to investigate whether the particle, weakly measured to have a non-vanishing weak value $(\Pi_B)_w$, will also have a chance to reach arm *E*. In other words, I want to investigate the possible appearance of the particle in arm *E given* that it has passed through arm *B*. To achieve this goal, it does not suffice to consider the weak measurements of $\Pi_B$ and $\Pi_E$ separately. Instead, I shall consider the *sequential weak measurement* of the projectors $\Pi_B$ and $\Pi_E$, a procedure that indeed tests for the presence of the particle in arm *E after* it has been present in arm *B*.

My extended argument is inspired by the paper [8] by Georgiev and Cohen, a paper that studies sequential weak measurements and the ensuing weak values (see also [9]). In its turn the paper [8] is an adaption and further elaboration of the weak value paradigm to the consistent-histories approach by Griffiths, Omnès, and Hartle and Gell-Mann (see, *e.g.*, [4] and references therein).

My simple argument, then, is the following. I want to establish beyond any doubt that a weak measurement of $\Pi_B$ results in the possibility for the particle to appear in arm *E*. The way to test whether this is the case is to make a (weak) measurement of the projector $\Pi_E$ *conditioned* on the particle having passed arm *B*, *i.e.*, conditioned on a positive outcome of a weak measurement of $\Pi_B$. To do that, following [8,9] I imagine attaching a von Neumann meter to the arm *B*, weakly measuring the projector $\Pi_B$ at a time $t_1$, and another von Neumann meter attached to the *E*-arm, weakly measuring the projector $\Pi_E$ at a later time $t_2$. As is shown in [8,9], by suitably combining the readings of the two meters, one will then be able to extract the sequential weak value $\{\Pi_B(t_1)\,\Pi_E(t_2)\}_w$ of the two-time correlation operator $\Pi_B(t_1)\,\Pi_E(t_2)$.[1]

Next, I rely on the physical interpretation of this weak value as an indicator of the presence of the particle in arm *E conditioned* on the particle's earlier appearance in arm *B*. Indeed, if this weak value is different from zero, it tells that the particle, which by weak measurement of $\Pi_B(t_1)$ has been found in the arm $B$ at the time $t_1$, will also (by a subsequent weak measurement of $\Pi_E(t_2)$) have a chance to be found in arm $E$ at the later time $t_2$. In a sense, therefore, this correlation weak value traces the particle one step further than what a weak value of $(\Pi_B)_w$ or $(\Pi_E)_w$ separately does, since they can, without further analysis, at most test for the particle passing arm $B$ independently of which way it takes after leaving that arm (respectively its presence in arm $E$ irrespectively of how it arrived there).

---

[1] This is very much an argument in principle. While the extraction of a single weak value, like $(\Pi_B)_w$, is first order in the weak measurement strength, the extraction of $\{\Pi_B(t_1)\,\Pi_E(t_2)\}_w$ is second order in the corresponding strengths. Whether this is at all possible experimentally is a different question.

To derive the expression for $\{\Pi_B(t_1)\,\Pi_E(t_2)\}_w$, one follows the evolution of the state of the particle through the nested MZI, including the effects of the weak measurements of $\Pi_B$ and $\Pi_E$. A straightforward calculation gives the result that $\{\Pi_B(t_1)\,\Pi_E(t_2)\}_w$ is non-zero. The conclusion, then, is clear: the particle that had a chance to pass arm $B$ has also a chance to pass arm $E$. The disturbance from the weak measurement of $\Pi_B$ so to speak "induces" the possibility for the particle to proceed through arm $E$ (and to ultimately reach the detector $D_2$), a situation that does not describe the undisturbed system.

This strengthens my previous arguments against Vaidman's conclusion: the passage of the particle through arm $B$ into arm $E$ is a result of the disturbance of the system, arising from the weak measurement of $\Pi_B$. The weak value $(\Pi_B)_w$ is not a *bona fide* property of the *undisturbed* system.

In conclusion, the use of weak sequential measurements furnishes a further argument against Vaidman's conclusions of particles appearing and disappearing in the nested MZI.

**Acknowledgement**

I am grateful to E. Cohen for valuable comments to an early draft of this paper.

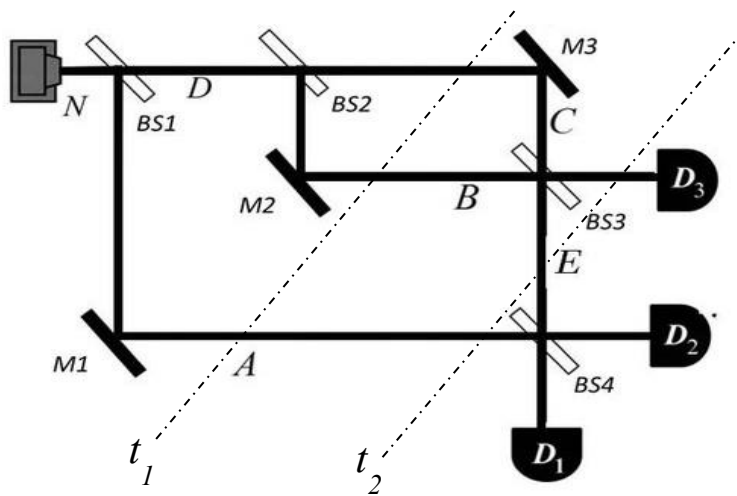

Figure. The nested MZI setup, exhibiting the notation for the arms (*N, D, etc.*), mirrors (*M1, etc.*), beamsplitters (*BS1, etc.*) and relevant times ($t_1$, $t_2$) used in the text (adopted from Vaidman [1] ),